%% file: main.tex
\documentclass[conference,final]{IEEEtran}
\IEEEoverridecommandlockouts
\usepackage{cite}
\usepackage{amsmath,amssymb,amsfonts}
\usepackage{algorithmic}
\usepackage{graphicx}
\usepackage{textcomp}
\usepackage{xcolor}
\usepackage{enumerate}
\usepackage{booktabs}
\usepackage{xspace}
\usepackage{multirow}
\usepackage{url}
\usepackage{algorithm}
\makeatletter
\newcommand{\removelatexerror}{\let\@latex@error\@gobble}
\makeatother

\makeatletter
\DeclareRobustCommand\onedot{\futurelet\@let@token\@onedot}
\def\@onedot{\ifx\@let@token.\else.\null\fi\xspace}

\def\eg{\emph{e.g}\onedot} 
\def\ie{\emph{i.e}\onedot} 
 
\def\etc{\emph{etc}\onedot} 
 
\def\etal{\emph{et al}\onedot}
\makeatother

\def\BibTeX{{\rm B\kern-.05em{\sc i\kern-.025em b}\kern-.08em
    T\kern-.1667em\lower.7ex\hbox{E}\kern-.125emX}}
\begin{document}

\title{{\huge MixDec Sampling: A Soft Link-based Sampling Method of Graph Neural Network for Recommendation}\\
\thanks{$\star$ Xiangjin Xie, Yuxin Chen, Ruipeng Wang and Xianli Zhang contributed equally to this research.}
\thanks{$\dagger$ Hai-Tao Zheng and Buyue Qian is the corresponding authors.}
}
\author{Xiangjin Xie$^{1\star}$, Yuxin Chen$^{2\star}$, Ruipeng Wang$^{3\star}$, Xianli Zhang$^{4\star}$, Shilei Cao$^2$, Kai Ouyang$^1$,\\
  Zihan Zhang$^5$, Hai-Tao Zheng$^{1,7\dagger}$, Buyue Qian$^{6\dagger}$, Hansen Zheng$^4$, Bo Hu$^2$, Chengxiang Zhuo$^2$, Zang Li$^2$\\
$^1$Shenzhen International Graduate School, Tsinghua University, $^2$Tencent PCG,\\
$^3$School of Mathematical Science, University of Electronic Science and Technology of China,\\
$^4$School of Computer Science and Technology, Xi’an Jiaotong University,\\ 
$^5$Beihang University, $^6$Beijing Chaoyang Hospital, Capital Medical University, $^7$Pengcheng Laboratory\\
$^1${\tt \{xxj20, oyk20\}@mails.tsinghua.edu.cn}, $^1${\tt zheng.haitao@sz.tsinghua.edu.cn},\\
$^2${\tt \{danikachen, eliasslcao, harryyfhu, felixzhuo, gavinzli\}@tencent.com},\\
$^3${\tt gavinwang@std.uestc.edu.cn}, $^4${\tt \{xlbryant, Hansen\}@stu.xjtu.edu.cn},\\
$^5${\tt zihanzhang@buaa.edu.cn}, $^6${\tt qianbuyue@bjcyh.com}
}

\maketitle

\begin{abstract}
Graph neural networks have been widely used in recent recommender systems, where negative sampling plays an important role.
Existing negative sampling methods restrict the relationship between nodes as either hard positive pairs or hard negative pairs. This leads to the loss of structural information, and lacks the mechanism to generate positive pairs for nodes with few neighbors.
To overcome limitations, we propose a novel soft link-based sampling method, namely MixDec Sampling, which consists of Mixup Sampling module and Decay Sampling module.
The Mixup Sampling augments node features by synthesizing new nodes and soft links, which provides sufficient number of samples for nodes with few neighbors.  
The Decay Sampling strengthens the digestion of graph structure information by generating soft links for node embedding learning.
To the best of our knowledge, we are the first to model sampling relationships between nodes by soft links in GNN-based recommender systems.
Extensive experiments demonstrate that the proposed MixDec Sampling can significantly and consistently improve the recommendation performance of
several representative GNN-based models on various recommendation benchmarks.

\end{abstract}

\section{Introduction}
\label{sec:intro}
\input{introduction}

\section{Preliminaries}
\label{sec:preliminaries}
\input{Preliminaries}

\section{Method}
\label{sec:method}

\input{Method}

\section{Experiments}
\label{sec:experiments}
\input{Experiments}

\section{Related Work}
\label{sec:Background}
\input{RelatedWork}

\section{Conclusion}
\label{sec:conclusion}
\input{conclusion}

\section{Acknowledge}
This research is supported by National Natural Science Foundation of China (Grant No.62276154 and 62011540405), Beijing Academy of Artificial Intelligence (BAAI), the Natural Science Foundation of Guangdong Province (Grant No. 2021A1515012640), Basic Research Fund of Shenzhen City (Grant No. JCYJ20210324120012033 and JCYJ20190813165003837), and Overseas Cooperation Research Fund of Tsinghua Shenzhen International Graduate School (Grant No. HW2021008).

\bibliographystyle{ieeetr}
\bibliography{ref}

\end{document}

%% file: introduction.tex
With the rapid increase of information on the Internet, recommender systems are widely used in e-commerce, social media, \textit{etc}. They provide\textbf{} users with a personalized flow of information based on their interaction history. Since the data of the recommender system naturally have a graph structure\cite{wu2020gnnsurvey}, a large number of recommendation methods that built based on graph neural network~(GNN) have emerged in recent years, such as GCN \cite{gcn},  GraphSAGE\cite{graphsage}, GAT\cite{gat}, Pinsage \cite{pinsage}, and LightGCN \cite{he2020lightgcn}.

The typical setup of the GNN-based recommendation method is as follows: 
1) firstly, a graph that reveals the user's interaction behavior is constructed, where each node is a user or an item; 2) subsequently, multiple propagation layers are applied to learn the user and item embeddings. Each propagation layer aggregates neighbor features of a current node to derive its embedding; 3) finally, the overall pipeline is supervised by a loss function that aims to pull embeddings of positive user-item pairs to be closer, while pushing embeddings of negative pairs away.

Negative sampling plays an important role in optimizing the affinity between user and item under the GNN-based methods.
Commonly, a uniform distribution is used for negative sampling \cite{he2020lightgcn, rendle2012bpr}.
To improve the effect, some negative sampling methods are recently proposed.
For example, PinSage \cite{pinsage} and DNS \cite{dns} focused on selecting hard negative samples for improving the ability of dealing with difficult samples. 
Generative adversarial network~(GAN) based methods, such as IRGAN \cite{wang2017irgan}, AdvIR \cite{park2019adversarial}, KBGAN \cite{cai2017kbgan}, generated negative samples through GAN.



\begin{figure*}
    \centering
    
    \includegraphics[width=\linewidth]{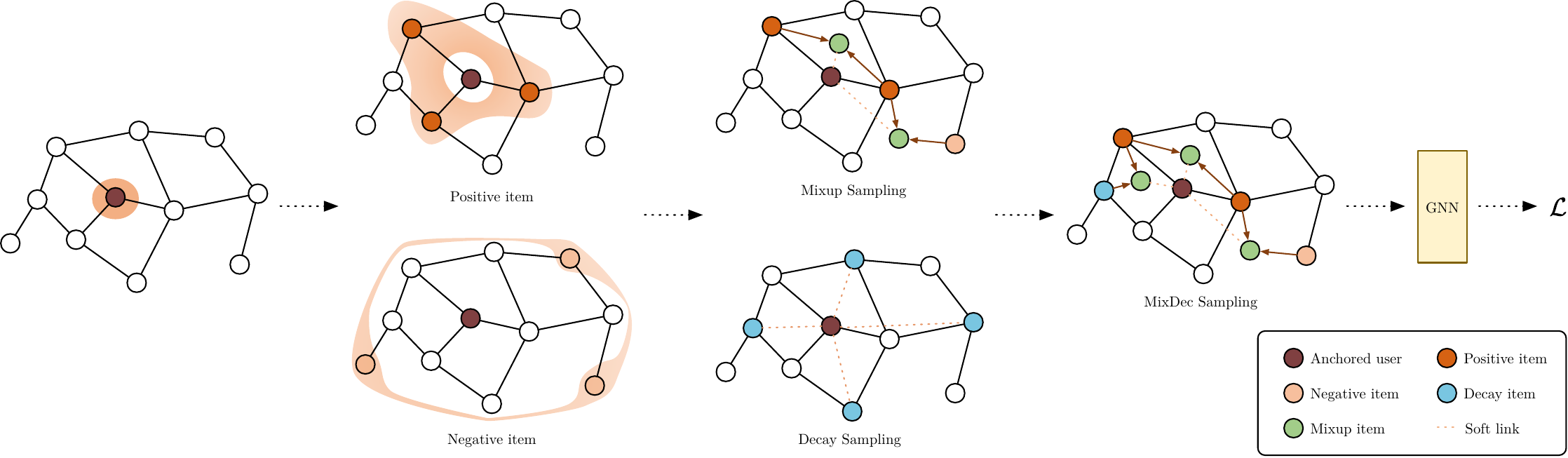}
    \caption{An overview of the MixDec sampling. MixDec consists of Mixup Sampling and Decay Sampling. 1) Based on sampled positive and negative items for the anchored user, Mixup Sampling linearly mixes $<$positive item, positive item$>$ pairs and $<$positive item, negative item$>$ pairs respectively to generate synthetic items. Then, the soft link is created between synthetic items and the user. 2) In Decay sampling, the decay item is the neighbor within the 
   l-hop for the anchored user. A soft link between $<$user, decay item$>$ will be built. In MixDec, Mixup Samples and Decay Samples will be integrated.}
    \label{fig:arc}
\end{figure*}

Despite the success of existing negative sampling methods, some limitations still exist. 
Firstly, in these methods, the relationship between two nodes is simply treated as either hard positive pairs or hard negative pairs.
However, nodes in a graph are typically connected by multi-hop neighbors, thus each pair of nodes naturally forms a soft proximity relationship.  Degenerating the original soft proximity relationship between nodes into a binary classification problem will obviously lead to the loss of structural information.
In addition, most of the existing negative sampling methods \cite{pinsage, dns, he2020lightgcn, wang2017irgan} mainly focused on sampling for constructing negative pairs, while lacking the generation of positive sample pairs. 
This may result in inadequate training of nodes with few neighbors.
 To overcome the above limitations, we propose a novel sampling method, namely \textbf{MixDec Sampling}, for GNN-based recommender systems~(see Fig. \ref{fig:arc} for an overview).
 \emph{MixDec Sampling} consists of Mixup Sampling module and Decay Sampling module. Firstly, the \textbf{Mixup Sampling} module augments node features of a graph by synthesizing new nodes and soft links.  It linearly mixes the features of positive samples and negative samples of each anchored node based on the Beta distribution, and fuses their links accordingly. Thus, nodes with few neighbors can be trained with more sufficient number of sample pairs. The \textbf{Decay Sampling} module generates soft links for each pair of nodes to strengthen the digestion of graph structure information in node embedding learning. The weights of links between nodes decay with their distance in a Breadth First Search (BFS) manner. In this way, the 
structural relationships between nodes are enriched. The Decay Sampling module cooperates with the Mixup Sampling module to boost the central idea of GNN---fusing features of neighbor nodes, meanwhile, utilizing the graph structure information.

To evaluate the effectiveness of our method, we conduct experiments by applying MixDec to representative GNN-based recommendation models (\eg GraphSAGE \cite{graphsage}, GCN \cite{gcn}, GAT \cite{gat}) and evaluate their performance on three recommendation benchmarks: Amazon-Book\footnote{Amazon-book, 2015. \url{http://jmcauley.ucsd.edu/data/amazon}}, Last-FM\footnote{Last-fm, 2015. \url{https://grouplens.org/datasets/hetrec-2011}}, and Yelp2018\footnote{Yelp, 2015. \url{https://www.yelp.com/dataset}}. 
The experimental results show that GNN-based models equipped with MixDec sampling method significantly and consistently outperform those equipped with the default negative sampling on various benchmarks.
In specific, we achieve average increases of 18.3\% for GraphSAGE and 21.1\% for GCN in terms of Mean Reciprocal Rank(MRR).

In summary, our work has the following contributions:
\begin{enumerate}
    \item To the best of our knowledge, we are the first to model relationships between nodes by soft links for sampling in GNN-based recommender systems, rather than 
    simply treats the relationship between nodes as either hard positive pairs or hard negative pairs. 
    \item We propose a novel sampling method, namely MixDec Sampling, for training GNN-based recommendation models. The proposed MixDec Sampling is able to boost the ability in learning meaningful embeddings from features and structural information of a user-item graph.
    \item Moreover, the proposed MixDec can be naturally plugged into existing GNN-based recommendation models.
    \item We empirically demonstrate that the proposed MixDec Sampling can significantly and consistently improve the recommendation quality of several popular GNN-based models on various recommendation benchmarks.
\end{enumerate}

%% file: Preliminaries.tex
In this section,  we introduce related concepts including GNN for recommendation, negative sampling, and Mixup \cite{mixup}.

\subsection{Graph Neural Networks for Recommendation}
 Recommendation is the most important technology in many e-commerce platforms, which has evolved from collaborative filtering to graph-based models. Graph-based recommendation represents all users and items by embedding and recommending items with maximum similarity score (by a inner product operation) for a given user. Here, we briefly describe the pipeline of GNN-based representation learning, including aggregation and optimization with negative sampling.

GNNs learn distributed vectors of nodes by leveraging node features and the graph structure. 
The neighborhood aggregation follows the ``message passing'' mechanism, which iteratively updates a node's embedding $h$ by aggregating the embeddings of its neighbors. Formally, the embedding $h_i^l$ of node $i$ in the $l$-th layer of GNN is defined as:
 \begin{equation} \label{agg}
     \small
     h_i^l= \sigma\left(\text{AGG}\left(h_{i}^{l-1}, h_{j}^{l-1} \mid j \in N_{(i)},W_l\right)\right),
 \end{equation}
where the \(\sigma\) is activation function, $W_l$ denotes the trainable weights at layer l, $N_{(i)}$ denotes all nodes adjacent to $i$, $\text{AGG}$ is an aggregation function implemented by specific GNN model (\eg GraphSAGE, GCN, GAT, \etc), and $h_i^0$ is typically initialized as the input node feature $v_i$.

\subsection{Negative Sampling}

Negative sampling \cite{negsamp} is firstly proposed to serve as a simplified version of Noise Contrastive Estimation\cite{NCE}, which is an efficient way to compute the partition function of an unnormalized distribution to accelerate the training of Word2Vec\cite{word2vec}. The GNN has different non-Euclidean encoder layers with the following negative sampling objective:

\begin{equation}\label{negsampling}
    \mathcal{L} = \log(\sigma (e_{v_i}^Te_{v_p}))+\sum^{c}_{j=1}\mathbb{E}_{v_j\sim P_n(v)}  \log(1-\sigma (e_{v_i}^Te_{v_j})),
\end{equation}
where $v_i$ is a node in the graph, $v_p$ is sampled from the positive distribution of node $v_i$, $v_j$ is sampled from the negative distribution of node $v_i$, $e$ represents the embedding of the node, $\sigma$ represents the sigmoid function, $c$ represents the number of negative samples for each positive sample pair. 

\subsection{Mixup}

\textbf{Mixup\cite{mixup}} is an simple yet effective data augmentation method that is originally proposed for image classification tasks. 
Mathematically, let $(x, y)$ denotes a sample of training data, where $x$ is the raw input samples and $y$ represents the one-hot label of $x$, the Mixup generates synthetic training samples $(\tilde{x}, \tilde{y})$  as follows:
\begin{equation}
\begin{split}
& \tilde{x}=\lambda x_{i}+(1-\lambda) x_{j}, \\
& \tilde{y}=\lambda y_{i}+(1-\lambda) y_{j}. \\
\end{split}
\end{equation}
It generates new samples by using linear interpolations to mix different images and their labels.

%% file: Method.tex
In this section, we propose MixDec Sampling, a soft link-based sampling method for GNN-based recommender systems. It can be plugged into existing GNN-based recommendation models, such as GraphSAGE, GCN and GAT.

An overview workflow of the MixDec Samping is illustrated in Fig. \ref{fig:arc}. Our MixDec Sampling boosts the GNN-based recommendation models in two aspects: (i) augmenting node features of a user-item graph by synthesizing new nodes and soft links, thus nodes with few neighbors can be trained with more sufficient number of samples; (ii) modeling the sampling relationships between nodes through soft links, where each link is weighted according to their distance in a Breadth First Search (BFS) manner. In this way, the richness of relationships between nodes is increased.

\subsection{\textbf{Mixup Sampling}}

\begin{figure}[t]
    \centering

    \includegraphics[width=.8\linewidth]{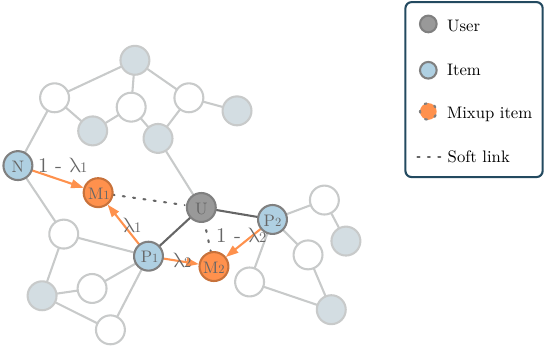}
    \caption{An illustration of the Mixup sampling. User \(U\) is the anchored node.  Due to directly connected to user \(U\),  \(P_1\) and \(P_2\) are positive samples. N is a negative sample. \(M_1\) and \(M_2\) are synthetic items, generated by mixing the feature of $<P_1, P_2>$ pair and $<P_1, N>$ pair according to the parameter \(\lambda_1\) and \(\lambda_2\), respectively. Then, we create a soft link with weight 1 between $<U, M_1>$ pair, and another soft link with weight $1-\lambda_2$ between $<U, M_2>$ pair. }
    \label{fig:mix_sam}
\end{figure}

To improve the embedding learning of nodes with few neighbors, we perform data augmentation on the node features in the graph by generating synthetic nodes and soft links. Mixup \cite{mixup} is a linear interpolation based data augmentation method with soft labels, which has been theoretically and empirically demonstrated the generalization and robustness.
Inspired by Mixup, we linearly mixes the features of positive and negative samples of each anchored node, and fuses their links accordingly.

For each anchored node in the graph, 
we set the weight of positively sampled nodes to one, and the weight of negatively sampled nodes to zero.
Then, we apply Mixup to positive and negative pairs to generate new pairs.
Specifically, each generated pair can be composed of two positive samples, or a positive sample and a negative sample.
The synthetic node and its link to the anchored node are obtained by Beta$(\alpha, \beta)$-based linear interpolation of the sampled nodes and weights of links, which is formalized as:

\begin{equation}\label{mix}
\begin{array}{c}
e_{s} = \lambda e_{i}+(1-\lambda) e_{j}, \\
w_{s} = \lambda w_{i}+(1-\lambda)w_{j}, \\
\end{array}
\end{equation}
where $e$ denotes the embedding of the node, $w$ represents the weight of the link between the anchored node and the another node. $s$ denotes the synthetic node, $i$ and $j$ denote the nodes obtained by positive and negative sampling of the anchored node and $\lambda \sim \text{Beta}(\alpha, \beta)$. An illustration of Mixup Sampling as shown in Fig. \ref{fig:mix_sam}.

The basic negative sampling loss $\mathcal{L}_{ns}$ is as follows:
\begin{equation}
\small
\mathcal{L}_{ns}=\frac{1}{N}\sum_{(v,n) \in\mathcal{O}} \log\left(\sigma\left ( e_v^T e_{n^+} \right )\right ) + \log\left(\sigma\left (-e_v^T e_{n^-}\right )\right ),
\end{equation}
where $N$ is the number of nodes in the graph, $ \mathcal{O} =\{(v,n) | (v,n^{+})\in \mathcal{R^{+}}, (v,n^{-})\in \mathcal{R^{-}}\} $ denotes the training set, \(\mathcal{R^{+}}\) indicates the observed~(positive) interactions between the anchored node $v$ and sampled node $n$, \(\mathcal{R^{-}}\) is the unobserved~(negative) interaction set. The loss of Mixup Sampling is then as follows:

\begin{equation}
\mathcal{L}_{m}=\frac{1}{N}\sum_{(v,s) \in\mathcal{O}_m} g\left(\sigma\left(e_v^T e_{s}\right), w_s\right),
\end{equation}
where $g$ is a loss function, $ \mathcal{O}_m$ denotes the Mixup training set,  which is the Mix interaction by (\ref{mix}).

For Mixup Sampling, the complete objective for optimization is defined as:
\begin{equation}
\begin{array}{c}

\mathcal{L}_{mix}=\mathcal{L}_{ns}+ \mathcal{L}_{m}.

\end{array}
\end{equation}


\subsection{\textbf{Decay Sampling}}

\begin{figure}[t]
    \centering

    \includegraphics[width=.8\linewidth]{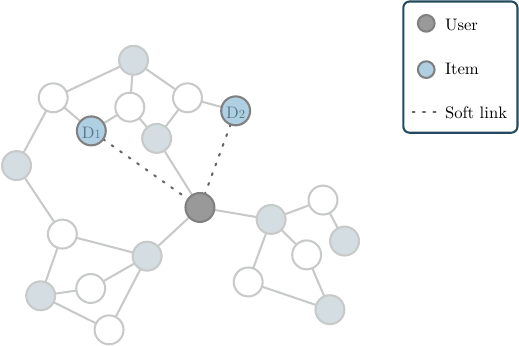}
    \caption{An illustration of the Decay sampling. \(D_1\) and \(D_2\) are decay items, which are the second-hop items found by anchored User \(U\) through BFS. Then a soft link is built between $<D_1,U>$ and $<D_2,U>$. Using soft link, the structural information of the graph is effectively preserved.}
    \label{fig:decay_sam}
\end{figure}

To preserve the structure information during the sampling process, we propose Decay Sampling that strengthens the digestion of graph structure information via a soft link mechanism.
The weights of links between nodes decay with their distance in a Breadth First Search~(BFS) manner.
An illustration of Decay Sampling is shown in Fig. \ref{fig:decay_sam}.


For each anchored node in the graph, we compute the decay weights of links between it and its neighbors within  \(l\)-hop based on BFS.
The weight of links is designed based on the number of reachable paths between the anchored node and its \(l\)-hop neighbors. 
The hop number \(l\) of BFS is the same as the number of aggregation layers in the GNN model. 
The weight of link \(w_d\) connecting the anchored node to the sampled node is defined as follows:

\begin{equation}\label{link}
    w_{d}=\rho+(1- \rho)\frac{r_d}{r_\text{max}},
\end{equation}
where $d$ is one of the non-positive sample of neighbors within \(l\)-hop reached by the anchored node $v$ on BFS paths, $w_{d}$ is the link weight between node $d$ and the anchored node, $\rho$ is used to map the weights of $w_{d}$ to $\left[\rho,1\right]$, $r_d$ is the number of pathways from anchored node $v$ to node $d$, and $r_\text{max}$ is the maximum value of all the \(r\) within \(l\)-hop by node $v$.

\begin{figure}[t]
\centering
\includegraphics[width=.8\linewidth]{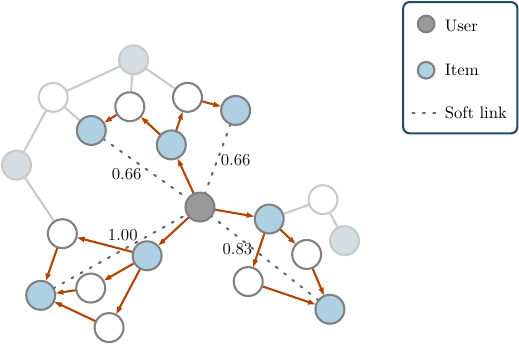}
\caption{An example of BFS-based decay weights of soft links. For the anchored user, items with direct interactions are regarded as positive items, and decay items are 3-hop neighbors. For these four decay items, the number of pathways from the user to the decay items, as obtained by BFS, is 1, 1, 3 and 2. Correspondingly, the decay weights of soft links are 0.66, 0.66, 1.00 and 0.83, when  the hyperparameter \(\rho\) is set to 0.5.} 
\label{fig:decayf} 
\end{figure}

We sort the sampled nodes by defined decay weights and intercept Top $k$. Meanwhile, the link weights of the rest nodes are set to 0. 
Fig. \ref{fig:decayf} shows an example of BFS-based dacay weight.

The Decay Sampling part loss is defined as follows:
 
\begin{equation}
\mathcal{L}_{d}= \frac{1}{N}\sum_{(v,d) \in\mathcal{O}_d} g\left(\sigma\left(e_v^T e_{d}\right), w_d\right),
\end{equation}
while the complete objective of Decay Sampling for optimization is as follows:

\begin{equation}\label{linkdecay}
\mathcal{L}_{dec}=\mathcal{L}_{ns}+ \mathcal{L}_{d},
\end{equation}
where $(v,d)\in \mathcal{O}_d$, $\mathcal{O}_d $ is the decay item set of each anchored node.

The time complexity of the BFS is $O(N ^{l+1})$.
Graphs in practical applications are not that dense, so the complexity is lower than $O( N^{l+1})$.
In particular, for a User-Item graph, suppose the number of user nodes is $N_u$, the number of item nodes is $N_i$ and $E$ is the number of edges in the graph, the time complexity $T$ is approximately:

\begin{equation}
    T=
    E\prod_{(i,u)\in V}^{\left \lfloor l/2\right \rfloor}  \frac{E}{N_i}\cdot\frac{E}{N_u}=\frac{E^l}{N_i^{\left\lfloor l/2\right\rfloor}N_u^{\left\lfloor l/2\right\rfloor}},
\end{equation}
where ${E}/{N_u}$ is the average number of user-item edges for each user node, ${E}/{N_i}$ is the average number of item-user edges for each item node and $l$ is odd. 

It is worth noting that the amount of data in the recommendation task is huge, \eg items may be viewed by millions or ten million users, which will cause the BFS to take too long to run.
Since random walks \cite{perozzi2014deepwalk} on graph has been used to estimate the reachability probability from anchored node to other nodes, the weight of the soft link in the above BFS can be approximated by random walk for large-scale graphs.
Specifically, we randomly walk $l$ steps starting from each anchored node. Then, $r_d$ in (\ref{linkdecay}) is the number of occurrences of sampled node $d$ in the walk sequences for the anchored node $v$.

\subsection{\textbf{MixDec Sampling}}

\begin{figure}[t]
    \centering

    \includegraphics[width=.8\linewidth]{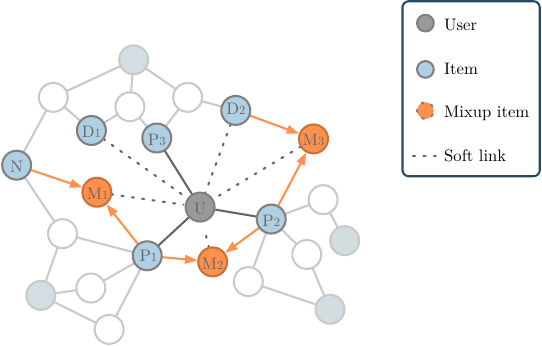}
    \caption{An illustration of the MixDec sampling. For example, in the figure, \(P_1\), \(P_2\), \(N\), \(M_1\) and \(M_2\) are the same as the Mixup sampling phase \ref{fig:mix_sam}, and \(D\) is the node searched by BFS in the Decay sampling phase. We synthesize \(M_3\) by mixing $<P_1, D>$ pair, and also build a soft link between $<U, M_3>$. MixDec sampling integrates Mixup sampling and Decay sampling, and the structure and feature information of the graph will be reserved.}
    \label{fig:mixdec_sam}
\end{figure}

MixDec Sampling is a joint sampling method of Mixup Sampling and Decay Sampling.
Mixup Sampling synthesizes the nodes and links, and then Decay Sampling preserves the graph's structure with soft links.
An illustration of MixDec Sampling is shown in Fig. \ref{fig:mixdec_sam}.
Formally, a GNN-based model is trained according to the following loss function:

\begin{equation}\label{loss}
\mathcal{L}=\mathcal{L}_{ns}+ \mathcal{L}_{m} + \mathcal{L}_{d},
\end{equation}
where $\mathcal{L}_{ns}$ is  basic negative sampling loss, $\mathcal{L}_{m}$ is the Mixup Sampling part loss, $\mathcal{L}_{d}$ is the Decay Sampling part loss, respectively.
Unlike Mixup Sampling,
each node pair of MixDec Sampling includes decay nodes, that is, node pairs of positive and positive, node pairs of positive and negative, node pairs of positive and decay. The MixDec Sampling is summarized in Algorithm \ref{alg}.

\subsection{\textbf{Loss function}}\label{other}

\begin{figure}[t]
  
  \renewcommand{\algorithmicrequire}{\textbf{Input:}}
  \renewcommand{\algorithmicensure}{\textbf{Output:}}
  \removelatexerror
  
  \begin{algorithm}[H]
    \caption{The train process with MixDec Sampling}\label{alg}
    \begin{algorithmic}[1]
        \REQUIRE Graph $G_{train}$, $G_{test}$, Number of Samples $c, c_m, c_d$. \\the training set $\mathcal{O}$, epochs.
        \STATE Initialize nodes feature $H^0$, Model $M_{\theta}$, Loss $\mathcal{L} = 0$.
        \FOR{$\mathit{each \; user}$ $\mathit{v}$ in $G_\text{train}$}
            \STATE Get the set $\mathcal{O}_d$ of the decay item and soft link $w_{d}$ of $\mathit{v}$  based on BFS and (\ref{link}).
        \ENDFOR
        \FOR {$\mathit{epoch}$ in $\mathit{epochs}$}
            \STATE Obtain node features $H^k$ by aggregation of (\ref{agg}).
            \FOR{$\mathit{each \; <user,item> }$ pair in $G_{train}$} 
  \STATE Sample $c$ nodes $\mathit{n}^-$ from $\mathcal{R}^-$, $c_m$ node $j$ from $\mathcal{O}_m$, $c_d$ nodes $d$ and soft link $w_d$ from $ \mathcal{O}_d$.
       \STATE Get synthetic nodes $s$, synthetic feature $H_s$ and soft link $w_s$ by Mix sampled nodes based on Mixup Sampling defined in (\ref{mix}).
            \ENDFOR
    \STATE Descending the gradients $\nabla_{\theta}\mathcal{L}$ by (\ref{loss}) and update $\theta$.
        \ENDFOR

    \end{algorithmic}
  \end{algorithm}
  
\end{figure}



We use Mean Absolute Error~(MAE) as the loss function $g$ of MixDec Sampling. Then, the Mixup Sampling part loss is formalized as:
\begin{equation}
\mathcal{L}_{m}=\frac{1}{N}\sum_{(v,s) \in\mathcal{O}_m} \left|\sigma\left(e_v^T e_{s}\right)-w_s\right|,
\end{equation}
and the Decay Sampling part loss is formalized as:
\begin{equation}
\mathcal{L}_{d}= \frac{1}{N}\sum_{(v,d) \in\mathcal{O}_d} \left|\sigma\left(e_v^T e_{d}\right)-w_d\right|.
\end{equation}

%

%% file: Experiments.tex
In this section,  we evaluate the performances of the proposed sampling method on three benchmark datasets with three representative GNN-based recommendation models. Furthermore, to analyze the MixDec Sampling method, we perform the hyperparameter study and the ablation study. Specifically, we intend to address the four research questions listed below:
\begin{itemize}
    \item \textbf{Q1.} Are there improvements in the effect of applying MixDec Sampling to representative GNN-based recommendation models? (See \ref{Q1}).
    \item \textbf{Q2.} Whether the proposed sampling method can continuously and steadily improve the performances when plugged
into existing GNN-based recommendation models? (See \ref{Q2}).
    \item \textbf{Q3.} How does MixDec Sampling method perform for nodes with few neighbors? (See \ref{Q3}).
    \item \textbf{Q4.} How do hyperparameters affect the performance of the proposed method? (See \ref{Q4}).
\end{itemize}
\subsection{Experimental Settings}
\subsubsection{\textbf{Dataset}}
We use three public benchmark datasets, Amazon-Book, Yelp2018, and Last-FM to evaluate our method.
Each dataset contains users, items, and interactions between them. To guarantee the quality of datasets, we only keep users and items with more than ten interactions.

\begin{table}[t]
\centering
\renewcommand\arraystretch{2} 
\caption{Statistics of the datasets.}
\label{tab:dataset}
\resizebox{\linewidth}{!}{
\begin{tabular}{c|cccc}

\hline
Dataset  & \#Users  & \#Items & \#Interactions & Density \\ \hline \hline
Amazon-Book   & 70,679 & 24,915 & 847,733    & 0.000048 \\ \hline
Yelp2018 & 45,919   & 45,538 &  1,185,068     & 0.000566 \\ \hline
Last-FM  & 23,566  & 48,123  &  3,034,796    &  0.002676\\ \hline
\end{tabular}}
\end{table}


\begin{table*}[t]
\centering
\renewcommand\arraystretch{2}
\caption{Performance Comparison Results.}
\label{result}

\resizebox{\textwidth}{!}{
\begin{tabular}{c|c|ccc|ccc|ccc}
\hline
\multirow{2}{*}{Evaluation Metrics} & \multirow{2}{*}{Dataset} & \multicolumn{3}{c|}{Amazon-Book} & \multicolumn{3}{c|}{Yelp2018} & \multicolumn{3}{c}{Last-FM} \\
                        &                      & GraphSAGE & GCN & GAT & GraphSAGE & GCN & GAT & GraphSAGE & GCN & GAT \\ \hline \hline
\multirow{5}{*}{MRR} &Uniform Negative Sampling   &0.1821 & 0.1480& 0.1652 & 0.1754 & 0.1682& 0.1872&  0.1627&  0.1409 &0.1834 \\
&Mixup Sampling & 0.1899 & 0.1551& 0.1722 & 0.1805 & 0.1837& \textbf{0.2018}&  0.1705&  0.1466 &0.1820 \\ 

 &Decay Sampling  & 0.1928 & 0.1747& 0.1757 & 0.1814 & \textbf{0.1842}& 0.1926&  \textbf{0.2353}&  \textbf{0.1892} &0.1895 \\ 

 &MixDec Sampling  & \textbf{0.1940} & \textbf{0.1770}& \textbf{0.1777}& 0.1824 & 0.1822 & 0.1972&  0.2280&  0.1840 &\textbf{0.1904} \\

 &\textbf{Improvement}  & \textbf{6.54\%} & \textbf{19.59\%}& \textbf{7.57\%}& \textbf{3.99\%} & \textbf{9.51\%}& \textbf{7.79\%}&  \textbf{44.62\%}&  \textbf{34.27\%}&\textbf{3.81\%} \\\hline
 \multirow{5}{*}{Hit@30} &Uniform Negative Sampling   &0.5535 &0.4956 & 0.5102 & 0.6192 & 0.5923& 0.6884&  0.5524& 0.4992 &0.5956 \\
&Mixup Sampling & 0.5520 & 0.5152 & 0.5450 & 0.5302 & \textbf{0.6370} & 0.7098& 0.5758&  0.5590&  0.5923  \\

 &Decay Sampling  & 0.5653 & 0.5295 & 0.5304 & 0.6307& 0.6334& 0.7057& \textbf{0.6091}&  \textbf{0.5642} & 0.6012 \\

 &MixDec Sampling  & \textbf{0.5720} & \textbf{0.5352} & \textbf{0.5574} & \textbf{0.6340} & 0.6320& \textbf{0.7132}&  0.6078& 0.5596&\textbf{0.6031}\\ 

 &\textbf{Improvement } & \textbf{3.34\%} & \textbf{7.99\%}&\textbf{ 9.25\%}& \textbf{2.39\%} & \textbf{7.54\%}& \textbf{3.60\%}& \textbf{10.26\%}&  \textbf{13.02\%} &\textbf{1.25\%} \\\hline
\end{tabular}%
}
\end{table*}

\begin{itemize}
    \item \textbf{Amazon-Book}\cite{amazon}: Amazon-Book is a widely used dataset for product recommendation.
    \item \textbf{Last-FM}\cite{last-fm}: This is the music listening dataset collected from Last-FM online music systems. We take the subset of the dataset where the timestamp is from Jan, 2015 to June, 2015. 
    \item \textbf{Yelp2018}\cite{yelp}: This dataset is adopted from the 2018 edition of the Yelp challenge. 
\end{itemize}

We summarize the statistics of the three datasets in Table~\ref{tab:dataset}. 
The table contains the number of  user nodes, item nodes, and interactions in graphs, with the density of graphs. We construct user-item bipartite graphs, the same as \cite{pinsage, wang2019kgat, huang2021mixgcf}.
For each dataset, we randomly select 80\% of each user's interaction history as the training set, and use the rest as the test set. From the training set, we randomly select 10\% of the interactions as the validation set to tune the hyperparameters. 
We use observed user-item interactions as positive examples and use Uniform Negative Sampling to sample unconsumed items for users.
On this basis, more soft link sample pairs are sampled using our Mixup Sampling, Decay Sampling, and MixDec Sampling methods, which can clearly demonstrate that the performance improvement comes from our sampling method.

\subsubsection{\textbf{Baseline Models}}

We use three representative GNN-based recommendation models, \ie, GraphSAGE, GCN, and GAT, as the base models for our experiments \cite{wu2020gnnsurvey}, respectively, and standardize on a three-layer structure.

\begin{itemize}
\item \textbf{GraphSage}~\cite{graphsage}
is a framework for inductive representation learning on large graphs. GraphSAGE is used to generate low-dimensional vector representations for nodes and is especially useful for graphs that have rich node attribute information.

\item \textbf{GCN}~\cite{gcn}
is a type of convolutional neural network that can work directly on graphs and take advantage of their structural information.

\item \textbf{GAT}~\cite{gat}
is a neural network architecture that leverages masked self-attentional layers to assign weights to the different nodes of the aggregation process.
\end{itemize}

\subsubsection{\textbf{Evaluation Metrics}}
To evaluate the performance for recommendation task, Hit@K \cite{hitk} and mean reciprocal ranking (MRR) \cite{mrr} serve as evaluation methodologies. 
\begin{itemize}
    \item \textbf{MRR} is a metric that evaluates the returned sorted list for recommendation tasks. MRR assigns different scores to different ranks of item v in the ranked list $R$ of the inner product of \(u\) and \(v\), which is defined as follows: 
    \begin{equation}\label{mrr}
        \text{MRR}=\frac{1}{|D_\text{candi}|} \sum_{(u_i,v_i)\in D_\text{candi}}\frac{1}{\text{rank}_{v_i}},
    \end{equation}
    where $\text{rank}_{v_i}$ is the rank of $v_i$ in the ranked list of user-item pair.
    \item \textbf{Hit@K} is a coarse granularity evaluation metric for recommendation task. Hit@K assigns same score to the first K of the ranked list $R$, which is defined as follows:
    \begin{equation}\label{hitk}
        \text{Hit@K}=\frac{\text{\#hit@K}}{|D_\text{candi}|}.
    \end{equation}
\end{itemize}
The $D_\text{candi}$ is the candidate set of items when predicting the favorite item for each user.
This set consists of one ground truth and 499 randomly selected items from which the user has not interacted.
In our experiment, we set $K$ as 30.

\subsubsection{\textbf{Implementation Details}}
All models are implemented in the PyTorch framework \cite{paszke2019pytorch}, and the Adam optimizer\cite{kingma2014adam} is adopted. Uniform Negative Sampling\cite{negsamp}, Mixup Sampling, Decay Sampling, and MixDec Sampling share an identical implementation of the aggregator, mini-batch iterator, and loss function to provide a fair comparison. For the other basic settings: the embedding dimension of the node is 128, the number of training epochs is 2,500, the number $c$ in Uniform Negative Sampling is 20 and $c_m$ and $c_d$ in MixDec Sampling is 5. Source code is relased in \url{ https://github.com/a2093930/MixDec-Sampling}.

\subsection{Results and Analysis}
We replace the Uniform Negative Sampling with our approaches (\ie, Mixup Sampling, Decay Sampling, and MixDec Sampling) on representative GNN-based recommendation models and evaluate their performance on three recommendation benchmarks.

\subsubsection{\textbf{Performance Comparison (for Q1)}} \label{Q1}
Benchmarks are shown in Table \ref{result} that includes GNN-based recommendation models and the proposed sampling approach implemented. The following are the findings:

\begin{itemize}
\item It's hardly surprising that the Uniform Negative Sampling approach from the GNN model performs the worst in all evaluations, since it simply interprets the relationship between nodes as either hard positive or hard negative pairs.
\item Our method is the most effective in terms of all metrics. Graph augmentation with synthetic nodes and links, along with soft link-based sampling, proves its efficacy in this scenario.
\item In most cases, Decay Sampling outperforms Mixup Sampling, demonstrating that it is crucial to maintain structural information during sampling.
\item High-density graphs benefit more from Decay Sampling's performance. Decay Sampling, in particular, improves MRR scores by up to 44.62\% and Hit@30 scores by up to 13.02\% on the Last-FM dataset. Mixup Sampling performs significantly better on graphs with low-density, such as the Amazon-Book dataset and the Yelp2018 dataset.
\end{itemize}

\subsubsection{\textbf{Generalization Ability (for Q2)}}\label{Q2}
Table \ref{result} shows that our strategy consistently outperforms Uniform Negative Sampling on all three GNN-based recommendation models, proving its generalizability.
In comparison to GraphSAGE and GAT, MixDec Sampling on GCN is far superior. This is due to the fact that GCN's preliminary performance on the most of results is the lowest.
Our strategy considerably improves the less effective GNN-based model.

\subsubsection{\textbf{Performance Analysis on Different Density Graphs (for Q3)}}\label{Q3}
In order to evaluate the performance of our method on nodes with few neighbors, we drop edges in the training set in multiple proportions to generate subgraphs with different degrees of sparseness.
We completed this experiment with GCN, we set the dropping ratio to be 0\%, 20\%, 50\%, and 70\% on Amazon-Books for Uniform Negative Sampling, Mixup Sampling and MixDec Sampling.
The results are presented in Table \ref{Px}. The observations are as followed:
 \begin{itemize}
     \item When the dropping ratio is 0\%, 20\% and 50\%, the improvement of Mixup Sampling compared to Uniform Negative Sampling gradually increases with the decrease of graph density.
     Mixup Sampling has more significant improvements in the case of fewer neighbors, which illustrates the effectiveness of Mixup Sampling on nodes with few neighbors.
     \item When the dropping ratio is less than 0.7, the improvement of MixDec Sampling relative to Uniform Negative Sampling gradually increases with the increase of graph density, while Mixup Sampling exhibits the opposite trend. This shows that Decay Sampling performs better in the density graph, which is also reflected in the performance comparison results in Table \ref{result}. Overall, MixDec Sampling still performs significantly better than Uniform Negative Sampling and Mixup Sampling with a large dropping ratio, such as 50\%.
     \item When the dropping ratio is set to 0.7, each user in the graph has an average of 3.59 neighbors. Due to the low density of the graph, the effective information is limited. Therefore, compared to the higher graph density, the improvement of Mixup Sampling and MixDec Sampling becomes less.
 \end{itemize}

\begin{table}[t]
\centering
\renewcommand\arraystretch{2} 
\caption{ Performance Results on Different Density Graphs}
\label{Px}
\resizebox{\linewidth}{!}{
\begin{tabular}{c|cccc}
\hline
Dropping Ratio & 70\% & 50\% & 20\%  & 0\% \\ \hline \hline
Uniform Negative Sampling& 0.0986  & 0.1145 &0.1304 &0.1480  \\ \hline
Mixup Sampling &0.1034&0.1221& 0.1382 &0.1551\\ 
\textbf{Improvement} &\textbf{4.86\%} & \textbf{6.63\%}& \textbf{5.98\%}&\textbf{4.79\%}\\ \hline
MixDec Sampling & 0.1041& 0.1302  &0.1517&0.1770\\ 
\textbf{Improvement}&  \textbf{5.57\%}& \textbf{13.71\%} & \textbf{16.33\%} &\textbf{19.59\%}\\ \hline
\end{tabular}
}
\end{table}



\subsubsection{\textbf{Efficiency Comparison}}

Take GCN as an example, the comparison results of time consumption for training 2,500 epochs of Uniform Negative Sampling, Mixup Sampling, Decay Sampling and MixDec Sampling are shown in Table \ref{TC}.
Uniform Negative Sampling is the fastest on all datasets, because our three samplings all have to sample negative items.
Among the two essential components of MixDec Sampling, Decay Sampling has the shorter time consumption. 
Due to including Uniform Negative Sampling, Mixup Sampling, and Decay Sampling, MixDec Sampling takes the longest time of all methods.
Overall, MixDec Sampling contributes no more than 30 percent to the total time consumption of Uniform Negative Sampling.
Thus, our method does not increase too much time consumption.


\begin{table}[t]
\centering
\renewcommand\arraystretch{2} 
\caption{ Efficiency Comparison}
\label{TC}
\resizebox{\linewidth}{!}{
\begin{tabular}{c|ccc}
\hline
Runtime (sec.) & Amazon-Books & Yelp-2018 &Last-FM  \\ \hline \hline
Uniform Negative Sampling& 5,500 & 7,863 &18,050 \\ \hline
Mixup Sampling &6,323 &9,214 & 20,160 \\ \hline
Decay Sampling &6,180 &8,124 & 19,183 \\ \hline
MixDec Sampling & 6,665 &9,652   &21,220\\ \hline

\end{tabular}
}
\end{table}
\subsection{\textbf{Parameter Study (for Q4)}}\label{Q4}

\subsubsection{\textbf{Impact of Sampling Distribution}}



\begin{figure}[t]
    \centering
    \includegraphics[width=1.0\linewidth]{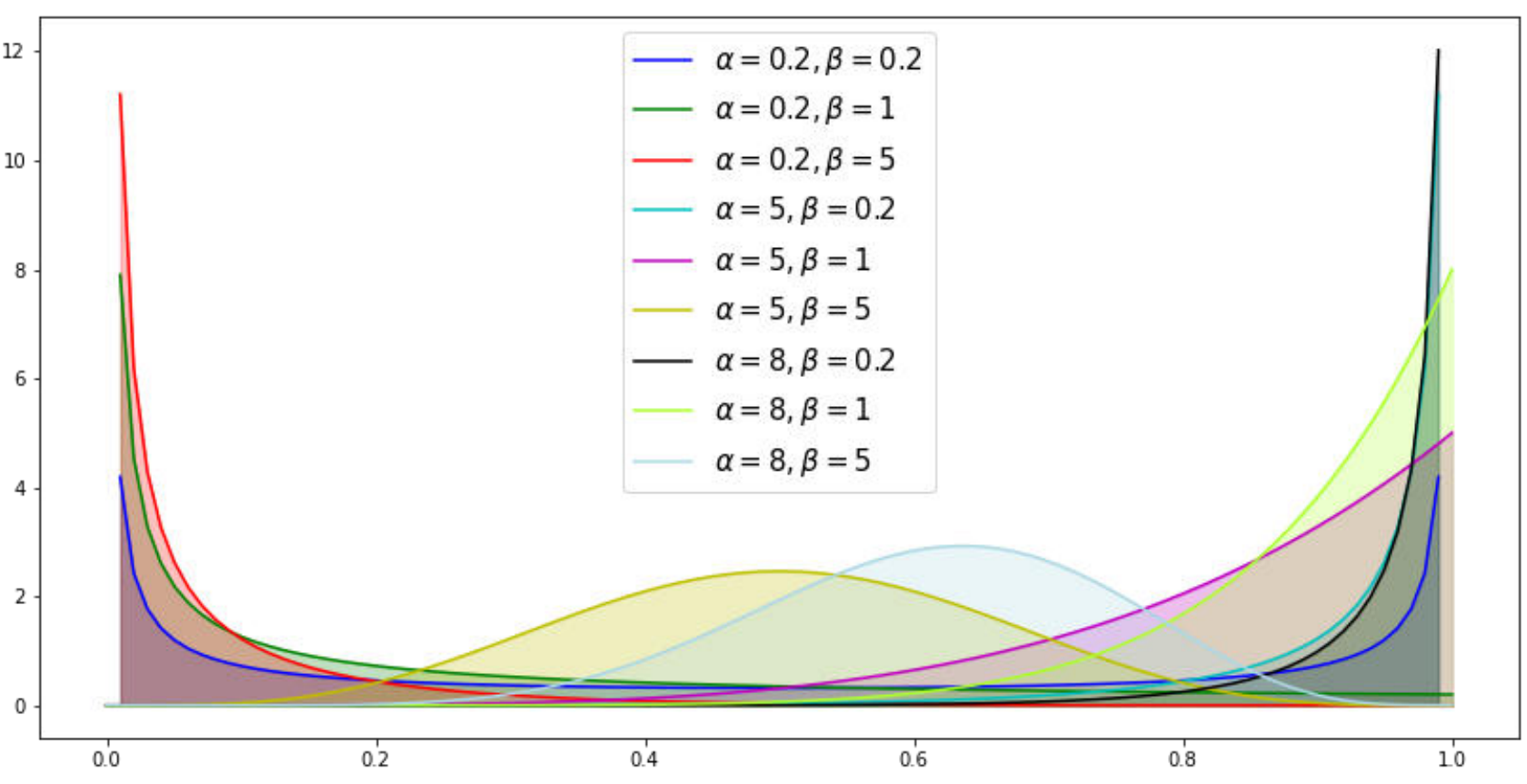}
    \caption{Distributions of Different $\text{Beta}(\alpha,\beta)$. The x-axis represents the value of $\lambda$, and the y-axis represents the corresponding probability density.}
    \label{fig:beta}
\end{figure}

We performed parameter sensitive experiments for \(\alpha\) and \(\beta\) of Beta distribution on the Amazon-Book dataset utilizing GCN to discover the optimal sampling distribution, and the results are shown in Table \ref{tab:lambda}. 
We choose $\alpha$, $\beta$ from $\{0.2,0.5, 0.8, 1, 5, 8\}$. Fig. \ref{fig:beta} describes multiple particular $\text{Beta}(\alpha,\beta)$ distributions to highlight how varied $\alpha$ and $\beta$ impacts the sampling distribution.

\begin{table}[t]
\centering
\renewcommand\arraystretch{2} 
\caption{ Impact of Beta Distribution}
\label{tab:lambda}
\resizebox{\linewidth}{!}{
\begin{tabular}{c|cccccc}
\hline
$\alpha \setminus \beta$& 0.2  & 0.5 & 0.8&1&5&8  \\ \hline \hline
0.2   & 0.1442 & \textbf{0.1570}& 0.1520 &0.1498&0.1419&0.1472   \\ \hline
0.5 &0.1455&0.1445 & 0.1401  &0.1400&0.1543&0.1487\\ \hline
0.8  & 0.1440  &0.1401  & 0.1451  &0.1512&0.1454&0.1502    \\ \hline
1 & 0.1362   & 0.1420  & 0.1518 &0.1504&0.1514&0.1560 \\ \hline
5 & 0.1503 & 0.1346 & 0.1348   &0.1369&0.1551&0.1550 \\ \hline
8  & 0.1466   & 0.1408  & 0.1375   &0.1357&0.1494&0.1522   \\ \hline
\end{tabular}
}
\end{table}

\begin{enumerate}[i.]
    \item (a) When both \(\alpha\) and \(\beta\) are less than 1 (\eg, $\alpha=0.2$, $\beta=0.2$), $\lambda \sim \text{Beta}(\alpha, \beta)$ is concentrated at 0 or 1. (b) When one of the values of \(\alpha\) and \(\beta\)  is greater than 1, \(\lambda\) changes monotonically (\eg, \(\lambda\) decreases monotonically when $\alpha=0.2$, $\beta=5$, and increases monotonically when $\alpha=8$ and $\beta=0.2$). (c) When both \(\alpha\) and \(\beta\)  are greater than 1 (\eg, $\alpha=5$, $\beta=5$), \(\lambda\) is centrally distributed between 0 and 1.
    
    \item As shown in Table \ref{tab:lambda}, $\alpha = 0.2$ and $\beta= 0.5$ achieves the best performance, while the corresponding value of $\lambda$ is around 0 and 1 according to Fig. \ref{fig:beta}. Similar results can be obtained with $\alpha = 0.2$ and $\beta= 0.8$. It illustrate that Mixup works better when $\lambda$ is around 0 or 1, suggesting that Mixup Sampling is more suited for producing small or large soft links.
    \item When $\alpha$ is 1 and $\beta$ is 0.2, most corresponding value of $\lambda$ is around 1, which has a poor performance.  This shows that Mixup Sampling is not suitable for all the $\lambda$ values close to 1.
\end{enumerate}


\subsubsection{\textbf{Impact of Decay Parameters}}

To study the contribution of $\rho$ and $k$ to Decay Sampling, we conduct the parameter sensitive experiment on the Amazon-Book dataset with GCN by choosing $\rho$ from $\{0,0.3, 0.5, 0.8, 1\}$ and $k$ from $\{100,200,300,400,500\}$. As shown in Table \ref{decay},  the best result is yielded when $\rho=0.5$ and $k=500$. The observations are as followed:
\begin{itemize}
    \item The parameter $\rho$ is used to project the weights of the soft link to $\left[\rho,1\right]$. If the parameter $\rho$ is set to a large value (\eg, 0.8, 1), the hierarchy between decay nodes will be obscured. If set $\rho$ to a small value (\eg, 0, 0.3), the decay node will not get sufficient positive information.  Experimental results show that the closer the value of $\rho$ is to 0.5, the better the result achieved. 
    \item The parameter $k$ indicates the number of decay items have been selected. When the value of $\rho$ is appropriate, the model performance improves with the increase of $k$~(Due to limited memory, we did not try larger values for $k$). Upon observation, the model performs optimally when $\rho$ is equal to 0.5 and $k$ is around 500. When $k$ and $\rho$ are both small (\eg $\rho=0$, $k=100$), the performance is the worst.
\end{itemize}

\begin{table}[t]
\centering
\renewcommand\arraystretch{2    }
\caption{Impact of Decay Parameters}
\label{decay}
\begin{tabular}{c|ccccc}
\hline
$\rho \setminus k$  & 100  & 200 & 300 & 400 &500 \\ \hline \hline
0   & 0.1173 &0.1304 & 0.1310  & 0.1269&0.1256 \\ \hline
0.3 & 0.1396 & 0.1405 &0.1517   & 0.1454&0.1480 \\ \hline
0.5  &0.1492   & 0.1511  & 0.1641 &0.1675& \textbf{0.1747}\\ \hline
0.8 & 0.1433   & 0.1452  & 0.1405& 0.1400&0.1442\\ \hline
1& 0.1450 & 0.1453  & 0.1363   & 0.1389 &0.1403 \\ \hline
\end{tabular}
\end{table}

%% file: RelatedWork.tex
In this section, we first provide a quick overview of GNNs and discussing their application in Recommendation tasks. Then, we introduce the current prevalent methods to negative sampling in GNNs, and their limitations. Finally, we introduce the data augmentation method of linear interpolation represented by Mixup \cite{mixup} and make a summary.

\subsection{GNNs for Recommendation}
In recent years, there has been a surge of research interest in developing varieties of GNNs, specialized deep learning architectures for dealing with graph-structured data. GNNs leverage the structure of the data as computational graph, allowing the information to propagate across the edges of graphs. A GNN usually consists of 1) graph convolution layers which extract local substructure features for individual nodes, and 2) a graph aggregation layer which aggregates node-level features into a graph-level feature vector.
Kipf \etal \cite{gcn} designed GCN by approximating localized 1-order spectral convolution.
Hamilton \etal \cite{graphsage} subsequently improved GCN which alleviate the receptive field expansion by sampling neighbors.
FastGCN \cite{fastgcn} further improved the sampling algorithm and adopts importance sampling in each layer.

Link prediction aims to find missing links or predict the likelihood of future links. Most of the contemporary approaches of link prediction focus on homogeneous networks where the object and the link are of single (same) types such as author collaboration networks. These networks comprise less information which may causes less accuracy for the prediction task. In heterogeneous networks, the underlying assumption of a single type of object and links does not hold good. Such networks contain different types of objects as well as links that carry more information compared to homogeneous networks and hence more fruitful to link prediction.

Huang \etal \cite{chen2005link} and Li \etal \cite{li2014recommendation} proposed approaches, where the recommender system (user–item recommendation) is represented as a bipartite graph, and employed basic link prediction approaches for the items recommendation. Sun \etal \cite{sun2009rankclus, sun2009ranking} coined the concept of heterogeneous information network (HIN) and subsequently meta path concept \cite{sun2011pathsim}, since then it becomes popular among researchers. Yang \etal \cite{yang2012predicting} proposed a new topological feature, namely multi-relational influence propagation to capture the correlation between different types of links and further incorporate temporal features to improve link prediction accuracy. Davis \etal \cite{davis2011multi} proposed a novel probabilistic framework. Their approach is based on the idea that the non-existing node pair forms a partial triad with their common neighbor, and their probabilistic weight is based on such triad census. Then the prediction score is computed for each link type by adding such weights.

\subsection{Negative Sampling}
Negative sampling is firstly proposed to speed up skip-gram training in Word2Vec \cite{word2vec}. Negative sampling has a significant impact on the performance of networks in GNN-based tasks, such as recommender systems and link prediction. As an example, for the link prediction and user recommendation scenarios, the recommendation model relies mostly on historical feedback from users to model their preferences. We learn the representation of users and items by feeding both positive and negative samples. In general, the interaction between users and items is treated as a set of implicit feedbacks. With implicit feedback, the database is not explicitly labeled, thus we have generally assumed all items that have interacted with the user are positive samples, and vice versa are negative samples. The approach to select negative samples can be broadly classified into heuristic negative sampling and model-based negative sampling.

Heuristic negative sampling focuses on sampling by setting some heuristic rules. Bayesian Personalized Ranking (BPR) \cite{rendle2012bpr} with negative sampling by uniform distribution over equal probabilities. The BPR algorithm has straightforward strategy and avoids introducing new biases in the sampling process, which is a widely used method. Popularity-biased Negative Sampling (PNS) \cite{chen2017sampling} applied the popularity of an item as the sampling weight, with the more popular the item, the more likely it is to be selected. The PNS increased the informativeness during sampling, but as its sampling distribution is calculated in advance, the sampling distribution does not change accordingly in the model training process \cite{rendle2014improving}. Therefore, the informativeness from negative sampling will decrease after several training sessions.

Model-based negative sampling aggregates the structural information of the model and obtains more high-quality negative samples. Dynamically Negative Sampling (DNS) \cite{zhang2013optimizing} dynamically changed the sampling distribution depending on the network, and the highest rated ones in the model were selected as negative samples for network training each time. MixGCF \cite{huang2021mixgcf} was a hop-wise sampling method which samples the representations of each hop among the negative. Besides, MixGCF applied the positive mixing to improve the quality of negative candidates. IRGAN \cite{wang2017irgan} used the idea of GAN for the first time in the field of information retrieval to perform negative sampling. The sampler performs as a generator and samples the negative to confuse the recommender. Then, \cite{ding2019reinforced, park2019adversarial} optimizing and improving the IRGAN in terms of efficiency and performance. To address the problem of inability to distinguish between the false negative and the hard negative when sampling, SRNS \cite{ding2020simplify} utilized the observed statistical features as a priori knowledge to separate the false negative and the hard negative, with a DNS-like structure for sampling to ensure the sampling quality and the robustness of the network.

\subsection{Mixup}

Interpolation-based data augmentation was proposed in Mixup \cite{mixup}. Mixup extends the training data by training a neural network on convex combinations of pairs of examples and their labels. Mixup has achieved relative success in many computer vision tasks. Mixup variants\cite{manifold,wordmixup,xie2022global} used interpolation in the hidden representation to capture higher-level information and obtain smoother decision boundaries. Recently, more researchers have focused on utilizing Mixup to improve the model’s performance in tasks with GNNs. GraphMixup\cite{wu2021graphmixup} presents a mixup-based framework for improving class-imbalanced node classification on graphs. MixGCF \cite{huang2021mixgcf} design the hop mixing technique to synthesize hard negatives. For graph classification, \cite{guo2021ifmixup} propose a simple input mixing schema for Mixup on graph, coined ifMixup, and they theoretically prove that, ifMixup guarantees that the mixed graphs are manifold intrusion free. G-Mixup augment graphs for graph classification by interpolating the generator (\ie, graphon) of different classes of graphs.

Mix sampling is inspired by mixup and negative sampling, converting nodal relations into soft relations and optimizing the model by linear interpolation and $\text{Beta}(\alpha,\beta)$ distribution.

%% file: Conclusion.tex
In this work, we proposed a novel soft link-based sampling method namely MixDec Sampling for GNN-based recommendation models.
Instead of being restricted to using hard positive pairs and hard negative pairs, MixDec Sampling provides a soft link to measure the proximity relationship between nodes, and the synthesized nodes also provide data augmentation for nodes with few neighbors.
Extensive experiments demonstrate the proposed MixDec Sampling can improve the recommendation performance of
several representative GNN-based models significantly and consistently on various recommendation benchmarks.
We hope this work would inspire the future soft link sampling method for GNN-based recommendation systems  for
efficient and effective utilizing graph information.